\documentclass[12pt]{iopart}
\usepackage{epsfig}
\usepackage{color}
\usepackage{bm}

\def\ketm#1{  \left\vert  #1   \right\rangle   }

\def\mem#1#2#3{  \left\langle #1 \left\vert  #2 \right\vert #3 \right\rangle   }
\def\rmem#1#2#3{  \left\langle #1 \left\vert \left\vert  #2
                  \right\vert \right\vert #3 \right\rangle   }
\begin{document}

%
%
%
%
\title[Polarization studies on the two--step radiative recombination]{Polarization
studies on the two--step radiative recombination of highly--charged,
heavy ions}

%
%
%
%
\author{A.~V.~Maiorova$^1$,
A.~Surzhykov$^{2,3}$, S.~Tashenov$^{4,5}$, V.~M.~Shabaev$^1$,
S.~Fritzsche$^{2,6}$, G.~Plunien$^7$, Th.~St\"ohlker$^{2,3}$}

%
%
%
%

\address{$^1$ Department of Physics, St.~Petersburg State
University, Ulianovskaya~1, Petrodvorets, St.~Petersburg 198504,
Russia}

\address{$^2$ GSI Helmholtzzentrum f\"ur Schwerionenforschung GmbH,
Planckstrasse 1, D--64291 Darmstadt, Germany}

\address{$^3$ Physikalisches Institut, Ruprecht--Karls--Universit\"at Heidelberg,
Philosophenweg 12, D--69120 Heidelberg, Germany}

\address{$^4$ Stockholm University, AlbaNova University Center, Atomic
Physics Division, SE--10691 Stockholm, Sweden}

\address{$^5$ Royal Institute of Technology (KTH),
AlbaNova University Center, Nuclear Physics Division, SE--10691
Stockholm, Sweden}

\address{$^6$ Frankfurt Institute for Advanced Studies (FIAS), Ruth--Moufang--Strasse
1, D--60438 Frankfurt am Main, Germany}

\address{$^7$ Institut f\"{u}r Theoretische Physik, TU Dresden, Mommsenstrasse
13, D--01062 Dresden, Germany}

%
%
%
%
\begin{abstract}

The radiative recombination of a free electron into an excited state
of a bare, high--$Z$ ion is studied, together with its subsequent
decay, within the framework of the density matrix theory and Dirac's
relativistic equation. Special attention is paid to the polarization
and angular correlations between the recombination and the decay
photons. In order to perform a systematic analysis of these
correlations the general expression for the double--differential
recombination cross section is obtained by making use of the
resonance approximation. Based on this expression, detailed
computations for the linear polarization of x--ray photons emitted
in the (e, 2$\gamma$) two--step recombination of uranium ions
U$^{92+}$ are carried out for a wide range of projectile energies.

\end{abstract}


\section{Introduction}

Radiative recombination (RR) is one of the basic processes that
occurs in many stellar and laboratory plasmas as well as in
collisions of heavy ions with electrons at ion storage rings and
electron beam ion traps (EBIT). In this process, a free (or
quasi--free) electron is captured into a bound state of an ion
under the simultaneous emission of a photon. Because of their
practical importance, detailed RR studies have been carried out
during the last two decades for many elements and for a wide range
of collision energies. At the GSI storage ring in Darmstadt, for
example, a large number of experiments have been done for the
electron capture into bare high--$Z$ ions, giving rise to
hydrogen--like ions after the recombination has taken place
\cite{Stohlker94,Stohlker95,Stohlker97,Stohlker99}. In the earlier
experiments, the total and angle--differential cross sections have
been measured hereby mainly for the ground--state capture and were
found in good agreement with computations based on Dirac's
equation
\cite{Eichler95,Shabaev00,Surzhikov01,Shabaev02,Eichler02,Klasnikov02,Klasnikov05,FrI05,Eichler07}.
Apart from the RR into the $1s_{1/2}$ ground state, most recent
studies have dealt also with the electron capture into the excited
ionic states which later decay under the emission of one (or
several) characteristic photons
\cite{Eichler98,Surzhikovprl,Surzhikov02}. Such a subsequent decay
is characterized (apart from the well known energies) by its
angular distribution and polarization of the emitted photons. Both
of these properties are closely related to the magnetic sublevel
population of the excited ion as it arises from the electron
capture. Several experiments have been carried out during last few
years in order to study the angular distribution and linear
polarization of the subsequent photons and, hence, enabled one to
derive the alignment of the residual ions. For the capture of an
electron into the $2p_{3/2}$ state of a bare uranium ion, for
instance, a strong alignment was found especially for the residual
ions, both by experiment \cite{Stohlker97} and in computations
\cite{Surzhikovprl}.

\medskip

In most experiments on the radiative decay cascades of high--$Z$
ions, that were performed so far, the emission of the first,
recombination photon remained unobserved. Although some insight
about the collisional dynamics and electronic structure of heavy
ions can be gained already from such an \textit{individual}
analysis of the characteristic radiation, more information is
obtained, if both, the recombination and decay photons are
measured in \textit{coincidence}. Moreover, such photon--photon
coincidence studies may have a significant impact also for the
development of novel experimental methods and techniques. It was
recently argued, for example, that they may help to determine the
polarization properties of heavy ions beams \cite{SuF03}, a
request which has been recently made by several groups.
Information about the ion polarization is required for studying,
for example, the parity non--conservation (PNC) effects in
highly--charged ions or in heavy--ion collisions
\cite{Labzowsky01,GSI}.

\medskip

Despite of the importance of coincidence RR experiments for the
forthcoming heavy ion research, little attention was paid up to
now to their \textit{theoretical} foundation. A first step towards
the theoretical description of the $(e, 2\gamma)$ RR process has
been done only recently by us in Ref.~\cite{Surzhikov02}. In that
paper, we have investigated the angle--angle correlations between
the recombination and the subsequent decay photons, assuming that
the polarization state of the photons remain unobserved. Owing to
the recent advances in x--ray polarization techniques
\cite{Stohlker03,Tashenov06}, however, a
\textit{polarization--resolved} analysis of the correlated photon
emission might become feasible in the next few years.

\medskip

In this contribution, we study here the polarization correlations
between the recombination and the subsequent decay photons in the
$(e, 2\gamma)$ radiative recombination of bare, high--$Z$ ions.
These correlations can be described most easily in the framework
of the density matrix theory, based on Dirac's relativistic
equation. However, before we shall present details from this
theory, we first summarize in Section~\ref{sec_geometry} the
geometry under which the photon--photon polarization correlations
are to be considered. In section \ref{sec_theory}, then, we make
use of the resonance approximation in order to derive the general
expression for the double--differential RR cross section which
depends on the emission angles and the polarization states of both
photons. Starting from this cross section, we perform in Section
\ref{sub_section_scenarious} a theoretical analysis for two
selected scenarios of possible x--x coincidence studies. First, we
shall discuss the \textit{angle--polarization} correlation in
which the linear polarization of the characteristic radiation is
explored, while the spin states of recombination photons remain
unobserved. \textit{Vice versa}, the angular distribution of the
characteristic decay photons, following the emission of linearly
polarized RR photons, is discussed as a second scenario, then
called the \textit{polarization--angle} case. While, of course,
the derived correlation functions can be applied to all
hydrogen--like ions, detailed computations have been carried out
for the electron capture into the $2p_{3/2}$ state of initially
bare uranium ion, and along with its subsequent Lyman--$\alpha_1$
($2p_{3/2} \to 1s_{1/2}$) decay. Results of our calculations are
presented in Section \ref{sec_results} and indicate strong
correlations between the angular and polarization properties of
the recombination and decay photons. Finally, a brief summary is
given in Section \ref{sec_summary}.

\medskip

Relativistic units $\hbar = m_{e}=c=1$ are used throughout the paper
unless stated otherwise.


\section{Geometry of the two-photon radiative recombination}
\label{sec_geometry}

\begin{figure}
\epsfbox{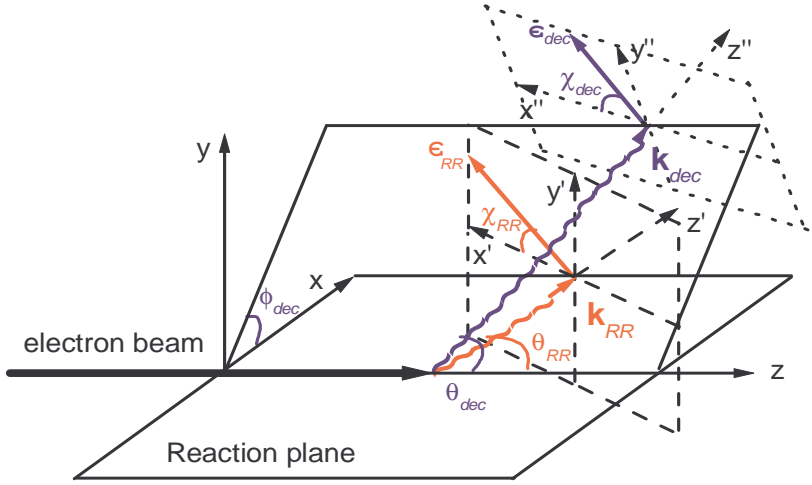} \caption{Geometry (in the ion rest frame) for the
radiative recombination of a free electron into an excited state of
a bare projectile ion, and followed by a subsequent photon decay.
The unit vectors of the linear polarization of the recombination and
decay photons are defined in the planes that are perpendicular to
the photon momenta $\bm{k}_1 \equiv \bm{k}_{RR}$ and $\bm{k}_2
\equiv \bm{k}_{dec}$, respectively.} \label{Fig1}
\end{figure}

In order to explore the polarization correlations in the two--step
radiative recombination of (finally) hydrogen--like ions, we shall
first agree about the geometry under which the emission of both, the
recombination and decay photons is observed. In the present work,
the angular-- and polarization--resolved properties of the photons
will be analyzed in the \textit{projectile frame} (i.e. the rest
frame of the ion). Since in this frame the only preferred direction
of the overall system is given by the electron momentum, here we
adopt the quantization axis ($z$--axis) along the direction of the
incoming electron (as \textit{seen} by the ion). Together with the
wave vector of the first photon $\bm{k}_1 \equiv \bm{k}_{RR}$, this
axis then defines also the reaction plane (\textit{x--z} plane).
Thus, only one polar angle $\theta_1 \equiv \theta_{RR}$ is required
to characterize the first, recombination photon, while the two
angles $(\theta_2, \phi_2)$ are used for describing the emission of
the subsequent decay photon (cf.~Fig.~\ref{Fig1}).

%
%

\medskip

For the theoretical analysis below we have to account for not only
the emission angles but also the \textit{linear} polarization
vectors $\bm{\epsilon}_{1}$ and $\bm{\epsilon}_{2}$ of the
recombination and decay photons. As usual, these vectors are defined
in the planes that are perpendicular to the photon momenta
$\bm{k}_1$ and $\bm{k}_2$ and are characterized by the angles
$\chi_1$ and $\chi_2$ with respect to the planes as spanned by the
quantization axis and the unit vectors $\hat{\bm{k}}_1$ and
$\hat{\bm{k}}_2$, respectively.


\section{Theoretical background}
\label{sec_theory}

%
%

\subsection{Resonance approximation}
\label{sub_section_resonance}

\begin{figure}
\epsfbox{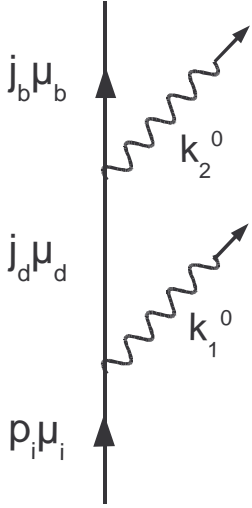} \caption{Feynman diagram for the two--step
radiative recombination of a free electron with a bare nucleus. In
this diagram, $p_i$ and $\mu_i$ are the four--momentum and spin
projection of the incoming electron, $\ketm{n_d j_d}$ and
$\ketm{n_b j_b}$ denote the intermediate and final (bound)
hydrogenic states, and $k_{1,2}$ represent the four-momenta of the
recombination and decay photons, respectively.} \label{Fig2}
\end{figure}

Having defined the geometry of the two--step radiative
recombination, we are prepared now to derive expression for the
differential cross section (DCS) of the process. The evaluation of
such cross sections is usually traced back to the RR transition
amplitude which, to zeroth order, is described by the diagram
shown in Fig.~\ref{Fig2}. In this diagram,
$p_{i}=(p^{0}_{i},\mathbf{p_{i}})$ and $\mu_{i}$ are the
asymptotic four--momentum and the spin projection of the incoming
electron, and $k_{1,2}=(k^{0}_{1,2},\mathbf{k_{1,2}})$ denote the
four-momenta of the first (recombination) and second (decay)
photon, respectively. By assuming that---for a given energy of the
incoming electrons---the x--ray detectors observe only those
photons which are emitted in course of (i) the electron capture
into some (excited) ionic state $\ketm{n_d j_d \mu_d}$ and (ii)
the $\ketm{n_d j_d \mu_d} \to \ketm{n_b j_b \mu_b}$ subsequent
decay to the ground state, we may restrict our theoretical
analysis to the \textit{resonance approximation} (cf.\
Ref.~\cite{Shabaev02} for further details). Within such an
approximation, the differential RR cross section reads as:
\begin{eqnarray}
   \label{dcs_general}
   d\sigma &=&
   \frac{(2\pi)^{4}}{v_{i}}
   \left|\sum_{\mu_d}
   \frac{
   \mem{n_b j_b \mu_b}{e\alpha^{\nu}A^{\ast}_{2,\nu}}{n_d j_d \mu_d}
   \mem{n_d j_d \mu_d}{e\alpha^{\mu}A^{\ast}_{1,\mu}}{p_i \mu_i}}
   {\varepsilon_{b}+k^{0}_{2}-\varepsilon_{d}+i\frac{\Gamma_{d}}{2}}\right|^{2}
   \nonumber \\
   &\times& (p^{0}_{i}-k^{0}_{2}-\varepsilon_{b})^{2}(\mathrm{k}_{2}^{0})^{2} \,
   \rm{d}k^0_2 \, \rm{d}\mathbf{\Omega}_1 \, \rm{d}\mathbf{\Omega}_2 \, ,
\end{eqnarray}
where $\varepsilon_{d}$ and $\Gamma_{d}$ are the energy and the
width of the intermediate (excited) state, $\varepsilon_{b}$ is
the ground--state energy, $v_{i}$ denotes the velocity of the
incident electron in the projectile frame and
$\alpha^{\nu}=(1,\mbox{\boldmath $\alpha$})$ with $\mbox{\boldmath
$\alpha$} = (\alpha_x,\,\alpha_y,\,\alpha_z)$ being the vector of
Dirac matrices. In Eq.~(\ref{dcs_general}), moreover, the
interaction of the electron with the radiation field is
characterized by the operator $ e\alpha^{\nu} A_{\nu}^{*} =
-e\mbox{\boldmath
   $\alpha$}\cdot \mathbf{A}^{*}$ where the vector
potential \cite{Berestetskii71}
\begin{eqnarray}
   \label{A_definition}
   \mathbf{A}(\mathbf{r}) = \frac{\mbox{\boldmath
   $\epsilon$}\exp(\mathrm{i}\mathbf{k}\cdot\mathbf{r})}{\sqrt{2k^{0}(2\pi)^{3}}} \, ,
\end{eqnarray}
describes the plane photon wave with polarization $\bm{\epsilon}$
and energy $k^{0}=|\mathbf{k}|=\omega$.

\medskip

Owing to the energy conservation
\begin{equation}
   \label{energy_conservation}
   k_{1}^{0}=p_{i}^{0}-\varepsilon_{b}-k_{2}^{0} \, ,
\end{equation}
only one of the photon energies can be varied independently. The
radiative recom\-bination cross section (\ref{dcs_general}) is
therefore differential in the angle and energy of the second
photon but only with regard to the angle of the first photon.
Moreover, if we assume the width $\Gamma_{d}$ of the intermediate
excited state to be small compared to a distance to other states,
we may extend the $k_{2}^{0}$ integration to the interval
$(-\infty,\infty)$. Then by making use of the identity
\begin{eqnarray}
   \int\limits_{-\infty}^{\infty}
   \frac{dk^{0}_{2}}{\left(\varepsilon_{b}+k^{0}_{2}-\varepsilon_{d}\right)^{2}+
   \frac{\Gamma_{d}^{2}}{4}}=\frac{2\pi}{\Gamma_{d}} \, ,
\end{eqnarray}
we can easily integrate Eq.~(\ref{dcs_general}) also over the
energy of the second (decay) photon and finally obtain the
double--differential RR cross section:
\begin{equation}
   \label{dcs_general_2}
   \fl
   d\sigma =
   \frac{(2\pi)^{4}}{v_{i}} \,
   \frac{2\pi}{\Gamma_{d}} \, \omega^{2}_{1}\omega^{2}_{2}
   \,
   \left|\sum_{\mu_d}
   \mem{n_b j_b \mu_b}{R^{+}_{2}}{n_d j_d \mu_d}
   \mem{n_d j_d \mu_d}{R^{+}_{1}}{p_i \mu_i}
   \right|^{2}
   \rm{d}\mathbf{\Omega}_1 \rm{d}\mathbf{\Omega}_2 \, ,
\end{equation}
and where, for the sake of brevity, we have introduced the notation
$R_{i} \equiv e \alpha^{\mu} A_{i, \mu}$ in order to denote the
electron--photon interaction operator.

%
%
%
%
%

%
%

\subsection{Alignment of excited ionic states}
\label{sub_section_alignment}

Equation~(\ref{dcs_general_2}) as derived in the previous Section
enables one to analyze the \textit{angular} and
\textit{polarization} correlations between the photons emitted in
the two--step radiative recombination of (initially) bare ions. In
order to perform such an analysis, it is useful to re--write the
doubly--differential RR cross section as:
\begin{eqnarray}
   \label{dcs_general_3}
   d\sigma &=&
   \frac{(2\pi)^{4}}{v_{i}}\frac{2\pi}{\Gamma_{d}}\omega^{2}_{1}\omega^{2}_{2}
   \sum_{\mu_{d} \mu^{'}_{d} \mu_b}
   \mem{n_b j_b \mu_b}{R^{+}_{2}}{n_d j_d \mu_d} \nonumber \\
   &\times&
   \mem{n_d j_d \mu'_d}{R^{}_{2}}{n_b j_b \mu_b}
   \mem{n_d j_d \mu_d}{\rho^{(d)}}{n_d j_d \mu'_d} \,
   \rm{d}\mathbf{\Omega}_1 \rm{d}\mathbf{\Omega}_2 \, ,
\end{eqnarray}
where we assume that incoming electrons are unpolarized and that the
spin state of the residual ion remains unobserved. In
Eq.~(\ref{dcs_general_3}), moreover, we have introduced the
\textit{density matrix} of the intermediate ionic state:
\begin{eqnarray}
   \label{intermediate_dm}
   \mem{n_d j_d \mu_d}{\rho^{(d)}}{n_d j_d \mu'_d} =
   \frac{1}{2}\, \sum\limits_{\mu_i}
   \mem{n_d j_d \mu_d}{R^{+}_{1}}{p_{i}\mu_{i}}
   \mem{p_{i}\mu_{i}}{R^{}_{1}}{n_d j_d \mu'_d} \, .
\end{eqnarray}
This density matrix describes the magnetic sublevel population of
the ion after the electron has been captured and the (first) RR
photon has left the system along the direction $\hat{n}_1 =
(\theta_1, \phi_1)$. Instead of using the density matrix
(\ref{intermediate_dm}), however, it is often more convenient to
represent the intermediate state of the ions in terms of the
so--called statistical tensors $\rho_{kq}^{(d)}$. Although, from a
mathematical viewpoint, the statistical tensors are equivalent to
the density matrix, they are constructed to represent the
spherical tensors of rank $k$ and component $q$ (cf.\
Refs.~\cite{Surzhikov02,Blum81,Balashov00} for further details):
\begin{eqnarray}
   \label{stat_tensors}
   \rho_{kq}^{(d)} =
   \sum_{\mu_{d} \mu'_{d}} (-1)^{j_{d}-\mu'_{d}}
   C_{j_{d}\mu_{d}, \,
   j_{d}-\mu'_{d}}^{kq}
   \mem{n_d j_d \mu_d}{\rho^{(d)}}{n_d j_d \mu'_d} \, ,
\end{eqnarray}
where $C_{j_{d}\mu_{d}, \, j_{d}-\mu'_{d}}^{kq}$ denote the
Clebsch--Gordan coefficients. Owing to the properties of these
coefficients, the tensor components $\rho_{kq}^{(d)}$ are nonzero
only for $0\leq k \leq 2j_{d}$ and $-k\leq q \leq k$.

\medskip

In the theory of atomic collisions, the statistical tensors
(\ref{stat_tensors}) are often re--normalized with respect to the
zero--rank tensor \cite{Blum81,Balashov00}
\begin{eqnarray}
   \label{alignment_def}
   A^{(d)}_{kq} = \frac{\rho_{kq}^{(d)}}{\rho_{00}^{(d)}} \, .
\end{eqnarray}
These reduced tensors (or \textit{alignment} parameters) are then
independent on the particular normalization of the ion density
matrix and are directly related to the relative population of the
individual substates $\ketm{n_d j_d \mu_d}$. By making use of these
alignment parameters we may finally write the differential RR cross
section as:
\begin{eqnarray}
   \label{dcs_general_4}
   d\sigma &=&
   \frac{(2\pi)^{4}}{v_{i}}\frac{2\pi}{\Gamma_{d}}\omega^{2}_{1}\omega^{2}_{2}
   \, \rho_{00}^{(d)} \,
   \sum_{\mu_{d}\mu^{'}_{d}\mu_{b}}\sum_{kq}
   (-1)^{j_{d}-\mu'_{d}}C_{j_{d}\mu_{d}, \,
   j_{d}-\mu'_{d}}^{kq}
   A_{kq}^{(d)} \nonumber \\
   &\times&
   \mem{n_b j_b \mu_b}{R^{+}_{2}}{n_d j_d \mu_d}
   \mem{n_d j_d \mu'_d}{R^{}_{2}}{n_b j_b \mu_b}
   \rm{d}\mathbf{\Omega}_1 \rm{d}\mathbf{\Omega}_2 \, .
\end{eqnarray}
This cross section still depends on the polarization states and
emission angles of both, the recombination and decay photons,
because of the dependence of the reduced statistical tensors
$A^{(d)}_{kq} = A^{(d)}_{kq}(\theta_1, \phi_1; {\bm \epsilon}_1)$
as well as the bound--bound transition amplitudes $\mem{n_d j_d
\mu'_d}{R^{}_{2}}{n_b j_b \mu_b}$ on the angular and polarization
properties of the emitted (first and second-step) photons.

%
%

\subsection{Evaluation of the free--bound and bound--bound transition
            amplitudes}
\label{sub_section_amplitudes}

As seen from Eqs.~(\ref{stat_tensors})--(\ref{dcs_general_4}), any
further analysis of the angular and polarization correlations
between the recombination and subsequent decay photons can be
traced back to the free--bound and bound--bound transition
amplitudes, $\mem{n_d j_d \mu_d}{-e \bm{\alpha}\cdot
\bm{A}}{p_{i}\mu_{i}}$ and $\mem{n_d j_d \mu_d}{-e
\bm{\alpha}\cdot \bm{A}}{n_b j_b \mu_b}$ respectively. Since the
evaluation of these matrix elements has been discussed in detail
elsewhere (cf.~Refs.~\cite{Eichler98, Surzhikov02}), here we just
restrict ourselves to a rather short account of the basic
relations. In particular, the evaluation of these matrix elements
is significantly simplified if their radial and spin--angular
parts are separated from each other. In order to perform such a
separation we have to employ the standard (two--component)
representation of Dirac's wavefunction and to \textit{decompose}
the electron--photon interaction operator (\ref{A_definition})
into its partial fields. Most naturally, this decomposition can be
carried out if we re--write the polarization vector of the photon
$\bm{\epsilon}$ in terms of two (linearly independent) basis
vectors $\bm{\epsilon}_{\lambda}$, with $\lambda = \pm 1$ being
the photon helicity (i.e. the spin projection on the direction of
propagation), and if we make use of the standard expansion
\cite{Rose57}:
\begin{eqnarray}
   \label{A_expansion}
   \fl
   {\bm \epsilon}_{\lambda}\exp(\mathrm{i}\mathbf{k}\cdot\mathbf{r})
   = \sqrt{2\pi} \, \sum_{J=1}^{\infty}\sum_{M=-J}^{M=J} \,
   \sum\limits_{p=0,1}
   \mathrm{i}^{J} \, (\mathrm{i} \lambda)^p \, \sqrt{2J+1} \mathbf{A}_{JM}^{p}(\mathbf{r})
   D^{J}_{M\lambda}(\mathbf{k}\rightarrow \mathbf{z}),
\end{eqnarray}
for the right-- ($\lambda = + 1$) and left--hand ($\lambda = - 1$)
circularly polarized light. In this expression, which has been
derived for an arbitrary choice of quantization ($z$--) axis,
$D^{J}_{M\lambda}(\mathbf{k}\rightarrow \mathbf{z})$ represents the
Wigner rotation matrix and $\mathbf{A}_{JM}^{p=0,1}(\mathbf{r})
\equiv \mathbf{A}_{JM}^{(m, {e})}$ are the usual magnetic and
electric multipole fields.

\medskip

Making use of the expansion (\ref{A_expansion}) and the
Wigner--Eckart theorem, we can now represent the bound--bound
transition amplitude
\begin{eqnarray}
   \label{amplitude_decomposition}
   \fl
   \mem{n_d j_d \mu_d}{-e \bm{\alpha}\cdot\bm{A}_\lambda}{n_b j_b \mu_b} &=&
   \frac{-e\sqrt{2\pi}}{\sqrt{2k^{0}(2\pi)^{3}}} \,
   \sum_{J=1}^{\infty} \, \sum_{M=-J}^{J} \, \sum\limits_{p=0,1}
   \mathrm{i}^{J} \, (\mathrm{i} \lambda)^p
   \sqrt{\frac{2J+1}{2j_d + 1}}
   \nonumber\\[0.2cm]
   &\times& D^{J}_{M\lambda}(\mathbf{k} \rightarrow
   \mathbf{z})
   \, C_{j_b \mu_b, \, J M}^{j_d \mu_d} \,
   \rmem{n_d j_d}{\bm{\alpha}\mathbf{A}_{J}^{p}}{n_b j_b}
\end{eqnarray}
in terms of its \textit{reduced} multipole matrix elements. These
reduced matrix elements can be easily splitted into their radial
and angular parts, and where the angular part can be evaluated
analytically by using the calculus of the irreducible tensor
operators. The radial part is in contrast represented by a
one--dimensional integral which has to be computed numerically.
For the details of these calculations we refer to
Refs.~\cite{Surzhikov02,Grant74,SuK05}.

\medskip

In contrast to the bound--bound transition amplitude, the
evaluation of the free--bound matrix elements $\mem{n_d j_d
\mu_d}{-e \bm{\alpha}\cdot \bm{A}}{p_{i}\mu_{i}}$ requires a
decomposition not only for the photon plane wave
(\ref{A_definition}) but also for the continuum electron wave
which still occurs with well defined asymptotic momentum $p_i$. As
discussed previously \cite{Eichler98}, the particular form of such
a decomposition depends on the choice of quantization axis. Using,
for example, the electron momentum $\bm{p}_i$ as the quantization
axis for the decomposition, the full expansion of the incoming
electron wave function is given by
\begin{eqnarray}
   \label{electron_wave_decompose}
   \ketm{p_{i}\mu_{i}}
   = \frac{1}{\sqrt{4\pi}} \, \frac{1}{\sqrt{p_{i}\varepsilon_{i}}}
   \, \sum_{\kappa}
   \mathrm{i}^{l}\exp(i\Delta_{\kappa}) \, \sqrt{2l+1} \,
   C_{l0, \,\frac{1}{2}\mu_{i}}^{j\mu_{i}} \,
   \ketm{\varepsilon_{i} \kappa \mu_{i}} \, ,
\end{eqnarray}
where $\Delta_{\kappa}$ is the Coulomb phase shift and
$\ketm{\varepsilon_{i} \kappa \mu_{i}}$ is the \textit{partial}
electron wave with the energy $\varepsilon_{i}=p^{0}_{i}$ and the
Dirac quantum number $\kappa=(-1)^{j+l+1/2}(j+1/2)$ determined by
angular momentum \emph{j} and parity of the state \emph{l}. This
expansion enables one to express the free--bound transition
amplitude
\begin{eqnarray}
   \mem{p_{i}\mu_{i}}{-e \bm{\alpha}\cdot\bm{A}}{n_d j_d \mu_d} &=&
   \frac{1}{\sqrt{4\pi}} \, \frac{1}{\sqrt{p_{i}\varepsilon_{i}}}
   \, \sum_{\kappa}
   (\mathrm{-i})^{l} \, \exp(-i \Delta_{\kappa}) \, \sqrt{2l+1}
   \nonumber \\
   &\times& C_{l0, \,\frac{1}{2}\mu_{i}}^{j\mu_{i}} \,
   \mem{\varepsilon_{i} \kappa \mu_{i}}{-e \bm{\alpha}\cdot\bm{A}}{n_d j_d \mu_d}
   \, ,
\end{eqnarray}
as a sum of partial amplitudes $\mem{\varepsilon_{i} \kappa
\mu_{i}}{-e \bm{\alpha}\cdot\bm{A}}{n_d j_d \mu_d}$ which, in
turn, can be evaluated by employing the photon wave decomposition
(cf. Eqs.~(\ref{A_expansion})--(\ref{amplitude_decomposition})).

%
%
\begin{figure}
\begin{center}
\includegraphics[width=0.9\textwidth]{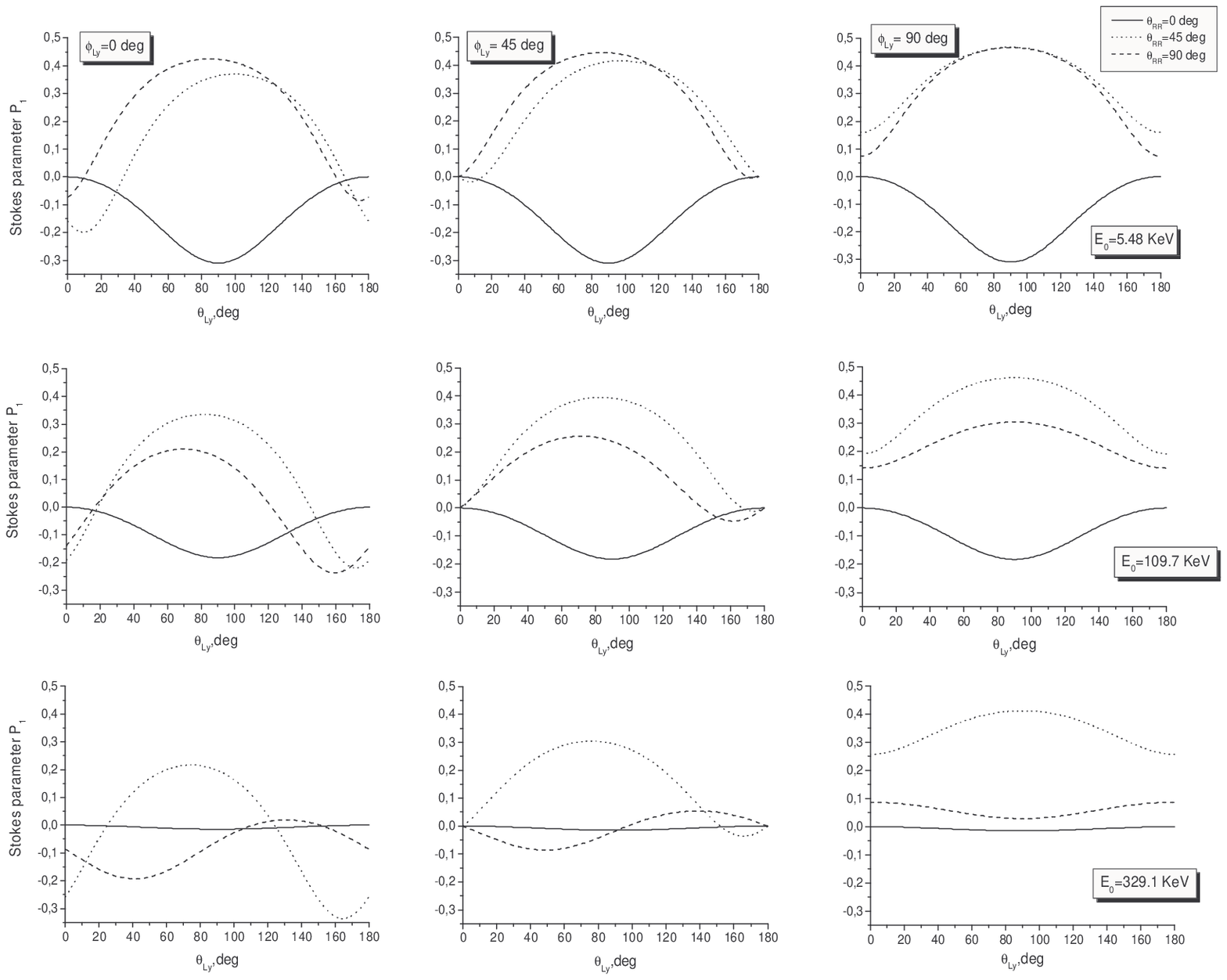}
\end{center}
\caption{Stokes parameter $P_{1}$ of the Lyman--$\alpha_1$ radiation
following the radiative recombination of a free electron into the
$2p_{3/2}$ state of the bare uranium projectile U$^{92+}$ with
energies $T_p$ = 10 MeV/u (upper row), 200 Mev/u (middle row) and
600 MeV/u (bottom row). These projectile energies correspond to the
kinetic energies of an incoming electron with $E_0$ = 5.48 keV,
109.7 keV and 329.1 keV in the ion--rest frame, respectively. The
polarization parameter is displayed in the projectile frame and for
the axial angles $\phi_2 \equiv \phi_{Ly}$ = 0$^\circ$ (left
column), 45$^\circ$ (middle column) and 90$^\circ$ (right column).}
\label{Fig3}
\end{figure}
%
%
%

%
%

\subsection{Polarization correlation studies}
\label{sub_section_scenarious}

Until now, we have discussed the evaluation of the
doubly--differential cross section (\ref{dcs_general_4}) for the
two--step radiative recombination. In the following, we shall apply
this RR cross section in order to study the angular and polarization
correlations between the two emitted photons. We will pay attention
especially to two scenarios for these studies. In Section
\ref{sub_sub_section_first_scenario}, we shall discuss the
polarization of the second, the \textit{decay} photon as measured in
coincidence with the recombination photon whose polarization remains
unobserved in this case. In contrast, the angular distribution of
the characteristic radiation, following the electron recombination
into an excited ionic state and together with the emission of a
linearly polarized recombination photon into some particular
direction will be obtained later in Section
\ref{sub_sub_section_second_scenario}.

%
%

\subsubsection{Angle--polarization RR studies:}
\label{sub_sub_section_first_scenario}

As said above, we shall first analyze the linear polarization of the
characteristic radiation by assuming that the polarization
properties of the recombination light are not resolved. With this
assumption in mind, the differential RR cross section reads as:
\begin{eqnarray}
   \label{dcs_general_scenario_1}
   d\sigma(\hat{n}_1 ; \, \hat{n}_2, \bm{\epsilon}_2) &=&
   \frac{(2\pi)^{4}}{v_{i}}\frac{2\pi}{\Gamma_{d}}\omega^{2}_{1}\omega^{2}_{2}
   \, \rho_{00}^{(d)} \,
   \sum\limits_{\lambda_1} \sum_{\mu_{d}\mu^{'}_{d}\mu_{b}}\sum_{kq} (-1)^{j_{d}-\mu'_{d}}C_{j_{d}\mu_{d}, \,
   j_{d}-\mu'_{d}}^{kq}
   A_{kq}^{(d)} \nonumber \\
   &\times&
   \mem{n_b j_b \mu_b}{R^{+}_{2}}{n_d j_d \mu_d}
   \mem{n_d j_d \mu'_d}{R^{}_{2}}{n_b j_b \mu_b}
   \rm{d}\mathbf{\Omega}_2 \, .
\end{eqnarray}
In this expression, we have performed the summation over the spin
states of the (first) recombination photons and have fixed its
emission angles $\hat{n}_1 = \left( \theta_1 , \phi_1 \right)$.
Despite this summation, Eq.~(\ref{dcs_general_scenario_1}) still
contains the \textit{complete} information about the polarization
and angular properties of the subsequent decay radiation which
follows the emission of the recombination radiation into a
particular direction $\hat{n}_1$.

\medskip

With the help of Eq.~(\ref{dcs_general_scenario_1}), we can evaluate
the \textit{linear} polarization of the decay photons. However,
before doing so we shall first agree about the parameters which are
used in order to characterize both, the degree as well as the
direction of such a polarization. From an experimental viewpoint,
the polarization of the emitted photons are most easily described in
terms of the so--called Stokes parameters which are determined by
the intensities of the light $I_\chi \sim d\sigma_\chi$ linearly
polarized under different angles $\chi$ with regard to the
\textit{reference} plane that is spanned by the quantization axis
(the beam direction) and the emitted photon momentum $\bm{k}$
(cf.~Fig.~\ref{Fig1}). For instance, while the parameter $P_1$
\begin{eqnarray}
   \label{P1}
   P_{1}(\hat{n}) = \frac{d\sigma_{0^{\circ}}-d\sigma_{90^{\circ}}}
   {d\sigma_{0^{\circ}} + d\sigma_{90^{\circ}}}
\end{eqnarray}
is derived from the intensities of light, polarized in parallel and
perpendicular to the reference plane, the parameter $P_2$ follows
from a similar ratio, taken at $\chi = 45^\circ$ and $\chi =
135^\circ$ respectively:
\begin{eqnarray}
   \label{P2}
   P_{2}(\hat{n}) = \frac{d\sigma_{45^{\circ}} - d\sigma_{135^{\circ}}}
   {d\sigma_{45^{\circ}} + d\sigma_{135^{\circ}}}
   \, .
\end{eqnarray}
As seen from these expressions, any polarization analysis of the
characteristic radiation requires the evaluation of the differential
cross section describing the emission of linearly polarized (decay)
photons under the angles $\hat{n}_2 = (\theta_2, \phi_2)$. Applying
the standard decomposition of the linear polarization vector in
terms of the circular polarization states \cite{Rose57}:
\begin{eqnarray}
   \label{linear_circular_expansion}
   \bm{\epsilon}_\chi =
   \frac{1}{\sqrt{2}}\sum_{\lambda = \pm 1} \exp(-\mathrm{i}\chi
   \lambda) \bm{\epsilon}_{\lambda},
\end{eqnarray}
such a differential cross section can be easily derived from
Eq.~(\ref{dcs_general_scenario_1}) and from the explicit form of the
electron--photon interaction operator $R_2$ as:
\begin{eqnarray}
   \label{dcs_general_scenario_2}
   \fl
   d\sigma(\hat{n}_1; \, \hat{n}_2, \bm{\epsilon}_{\chi_{_2}}) &=&
   \frac{(2\pi)^{4}}{v_{i}}\frac{4\pi^{2} \alpha}{\Gamma_{d}}\omega^{2}_{1}\omega^{2}_{2}
   \, \rho_{00}^{(d)} \,
   \sum\limits_{\lambda_1} \sum\limits_{\lambda_2 \lambda'_2}
   \sum_{\mu_{d}\mu^{'}_{d}\mu_{b}}\sum_{kq} \exp(\mathrm{i} {\chi}_2 (\lambda_2 -
   \lambda'_2))
   (-1)^{j_{d}-\mu'_{d}}    \nonumber \\
   & \fl\times& \fl A_{kq}^{(d)} C_{j_{d}\mu_{d}, \, j_{d}-\mu'_{d}}^{kq}
   \mem{n_b j_b \mu_b}{\bm{\alpha}\cdot\bm{A}^*_{\lambda_2}}{n_d j_d \mu_d}
   \mem{n_d j_d \mu'_d}{\bm{\alpha}\cdot\bm{A}_{\lambda'_2}}{n_b j_b \mu_b}
   \rm{d}\mathbf{\Omega}_2 \, .
\end{eqnarray}
Together with  Eq.~(\ref{amplitude_decomposition}), this expression
enables us to calculate the Stokes parameters (\ref{P1})--(\ref{P2})
of the subsequent decay photons. Apart from the emission angle
$\hat{n}_2 = (\theta_2, \phi_2)$ the polarization Stokes parameters
will depend also on the direction $\hat{n}_1 = (\theta_1, \phi_1)$
of the recombination photons, providing thus an opportunity to
investigate \textit{angle--polarization} correlations in the
two--step radiative recombination.

%
%
\begin{figure}
\begin{center}
\includegraphics[width=0.6\textwidth]{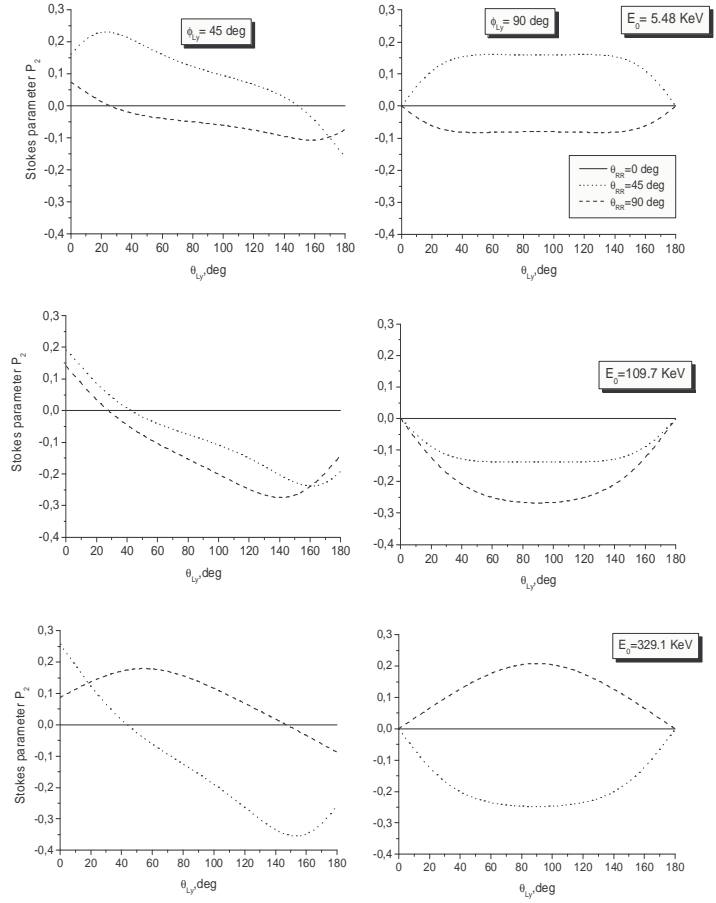}
\end{center}
\caption{ Stokes parameter $P_{2}$ of the Lyman--$\alpha_1$
radiation following the radiative recombination of a free electron
into the $2p_{3/2}$ state of the bare uranium projectile U$^{92+}$
with energies $T_p$ = 10 MeV/u (upper row), 200 Mev/u (middle row)
and 600 MeV/u (bottom row). These projectile energies correspond to
the kinetic energies of an incoming electron with $E_0$ = 5.48 keV,
109.7 keV and, respectively, 329.1 keV in the ion--rest frame. The
polarization parameter is displayed in the projectile frame and for
the axial angles $\phi_2 \equiv \phi_{Ly}$ = 45$^\circ$ (left
column), and 90$^\circ$ (right column). For the emission of the
decay photon within the reaction plane ($\phi_2$ = 0$^\circ$), the
Stokes parameter $P_2$ is zero.} \label{Fig4}
\end{figure}
%
%
%

%
%

\subsubsection{Polarization--angle RR studies:}
\label{sub_sub_section_second_scenario}

In our second case, we consider the angular distribution of the
characteristic photons with unobserved polarization that follow the
emission of (linearly polarized) recombination photons in some given
direction $\hat{n}_1 = (\theta_1, \phi_1)$. This angular
distribution
\begin{eqnarray}
   \label{dcs_general_scenario_3}
   \fl
   d\sigma(\hat{n}_1, \bm{\epsilon}_{\chi_{_1}}; \, \hat{n}_2) &=&
   \frac{(2\pi)^{4}}{v_{i}}\frac{2\pi}{\Gamma_{d}}\omega^{2}_{1}\omega^{2}_{2}
   \, \rho_{00}^{(d)} \, \sum\limits_{\lambda_2} \sum_{\mu_{d}\mu^{'}_{d}\mu_{b}}\sum_{kq}
   (-1)^{j_{d}-\mu'_{d}}
   \nonumber \\
   &\times& C_{j_{d}\mu_{d},\,
   j_{d}-\mu'_{d}}^{kq}
   A_{kq}^{(d)}
   \mem{n_b j_b \mu_b}{R^{+}_{2}}{n_d j_d \mu_d}
   \mem{n_d j_d \mu'_d}{R^{}_{2}}{n_b j_b \mu_b}
   \rm{d}\mathbf{\Omega}_2 \,
\end{eqnarray}
can be obtained from the general formula (\ref{dcs_general_4})
upon summation over the spin states of the subsequent photons and
by fixing the emission angle and the linear polarization angle
$\chi_1$ of the recombination light. In
Eq.~(\ref{dcs_general_scenario_3}), the dependence on the angle
and polarization properties of the first (recombination) photon
arises from the reduced statistical tensors $A_{kq}^{(d)} =
\rho^{(d)}_{kq}/\rho^{(d)}_{00}$ where
\begin{eqnarray}
   \label{rhokq_dependence}
   \fl
   \rho^{(d)}_{kq} \equiv
   \rho_{kq}^{(d)}(\theta_1, \phi_1, ; {\bm \epsilon}_{\chi_{_1}})
   &=& \pi\alpha\sum_{\mu_{d} \mu'_{d} \mu_i} \, \sum\limits_{\lambda_1 \lambda'_1}
   \, (-1)^{j_{d}-\mu'_{d}}
   \, \exp(\mathrm{i} \chi_1 (\lambda_1 - \lambda'_1)) \,
   C_{j_{d}\mu_{d},\,
   j_{d}-\mu'_{d}}^{kq} \nonumber \\
   & \fl\times& \fl \mem{n_d j_d \mu_d}{\bm{\alpha}\cdot\bm{A}^*_{\lambda_1}(\theta_1, \phi_1)}{p_{i}\mu_{i}}
   \mem{p_{i}\mu_{i}}{\bm{\alpha}\cdot\bm{A}_{\lambda'_1}(\theta_1, \phi_1)}{n_d j_d
   \mu'_d} \, .
\end{eqnarray}
By evaluating these statistical tensors and by employing them in
Eq.~(\ref{dcs_general_scenario_3}) we are able to investigate the
\textit{polarization--angular} correlations in the two--step
radiative recombination.


\section{Results and discussion}
\label{sec_results}

In the previous Sections we have derived the general formulas for
the differential radiative recombination cross section
(\ref{dcs_general_4}) as well as for the angle--polarization
(\ref{dcs_general_scenario_1}) and polarization--angle
(\ref{dcs_general_scenario_3}) correlation functions. While these
expressions can be applied of course to all hydrogenic states, a
more detailed analysis is performed in this work for the electron
capture into the $2p_{3/2}$ state of (initially) bare uranium ions
U$^{92+}$ and its subsequent Lyman--$\alpha_1$ ($2p_{3/2} \to
1s_{1/2}$) radiative decay. The angular and polarization properties
of the subsequent characteristic photon emission, as measured in
coincidence with the recombination light, will be in the focus of
forthcoming experiments at the GSI and FAIR facilities in the next
few years.

\medskip

Let us start our theoretical analysis of the angle and
polarization correlations in the two--step RR of bare uranium ions
from the computation of the Stokes parameters $P_{1}$ and $P_{2}$
for the Lyman--$\alpha_1$ photons. By following an
``experimentally realistic'' scenario as discussed in Section
\ref{sub_sub_section_first_scenario}, we here assume that
characteristic photons are measured in coincidence with the
recombination radiation whose polarization properties remain
however unobserved. For such a scenario, the polarization
parameters will depend on both, the nuclear charge $Z$ and the
projectile energy $T_p$ as well as on the angles under which the
recombination and decay photons are observed. In Figures
\ref{Fig3} and \ref{Fig4}, we display the Stokes parameters $P_1$
and $P_2$ as functions of the emission angle $\theta_2 \equiv
\theta_{Ly}$ (of the decay photon), and calculated for different
angles $\theta_1 = \theta_{RR}$ = 0$^\circ$, 45$^\circ$ and
90$^\circ$ for the emission of the recombination photon with
respect to the beam direction. In addition, we show these
(angular) distributions of the polarization parameters also for
three observation planes that are tilted by $\phi_2$ = 0$^\circ$,
45$^\circ$ and 90$^\circ$ with regard to the reaction plane, and
for the three projectile energies $T_p$ = 10, 200 and 600 MeV/u,
respectively. At these energies, the \textit{forward}
($\theta_{RR} = 0^\circ$) emission of the recombination photon
results in a parameter $P_1$ which is symmetric around
$\theta_{Ly} = 90^\circ$ and does not depend on the axial angle
$\phi_{Ly}$, while the Stokes parameter $P_2$ vanishes completely.
Such a behaviour of polarization parameters is well expected since
a recombination photon emission in forward (or backward) direction
does not break the axial symmetry for the intermediate ``excited
ion plus photon'' system and, hence, leads to a \textit{diagonal}
density matrix (\ref{intermediate_dm}) in this case. For this
reason, the magnetic sublevel population of the $2p_{3/2}$ excited
state can be described by a single \textit{non--zero} alignment
parameter $A_{20}^{(d)}$ (with $A_{2q}^{(d)} = 0$ for $q \ne 0$)
and this, in turn, results in the symmetric angular distribution
of the first Stokes parameter $P_1$ and a vanishing second
parameter $P_2$ (see Ref.~\cite{Balashov00} for further details).

\medskip

Of course, the symmetry of the intermediate system is broken in all
cases if the recombination photon is observed under an angle
$\theta_{RR} \ne 0^\circ$ and $\ne 180^\circ$. Then, the
polarization of the characteristic Lyman--$\alpha$ radiation is
described by non--zero parameters $P_1$ and $P_2$ which are
dependent on the axial angle $\phi_{RR}$ and asymmetric with respect
to the angle $\theta_{Ly} = 90^\circ$. Only if the characteristic
Lyman--$\alpha_1$ photons are measured perpendicular to the reaction
plane ($\phi_{Ly} = 90^\circ$), a symmetric distribution around
$\theta_{Ly} = 90^\circ$ is again restored as can be seen from
Eq.~(\ref{dcs_general_scenario_1}). For such a perpendicular
($\theta_{Ly} = 90^\circ$, $\phi_{Ly} = 90^\circ$) geometry, one may
indeed observe a rather strong linear polarization of the
Lyman--$\alpha_1$ line, especially if the recombination photon is
emitted under the angle $\theta_{RR} = 45^\circ$ with respect to the
beam direction. As seen from Fig.~\ref{Fig3}, for these angles the
first Stokes parameter slightly decreases from $P_1$ = 0.47 to
$0.41$ if the projectile energy is increased from $T_p$ = 10 MeV/u
to 600 MeV/u.

\medskip

Until now we analyzed the linear polarization of the characteristic
Lyman--$\alpha_1$ ($2p_{3/2} \to 1s_{1/2}$) line as measured in
coincidence with the recombination photons. In this analysis, we
assumed that the spin states of the recombination radiation remain
unobserved. In the following, we discuss the ``inverse'' situation
when the (angular) properties of the subsequent decay are observed
for the case of a well--defined polarization state of the
recombination radiation. Again, we restrict ourselves to the
Lyman--$\alpha_1$ line whose angular distribution can be obtained
from the general expression (\ref{dcs_general_scenario_3}) in the
form:
\begin{eqnarray}
   \label{angular_Lyman}
   \frac{d\sigma(\hat{n}_1, \bm{\epsilon}_{\chi_{_1}}; \,
   \hat{n}_2)}{d\Omega_2} &=& \frac{\sigma_0}{4 \pi}
   \, \left( 1 + \sqrt{\frac{\pi}{5}}
   \sum\limits_q Y_{2q}(\theta_2, \phi_2) \,
   A^{(d)}_{2q}(\hat{n}_1; {\bm \epsilon}_{\chi_{_1}}) \, f_2^{3/2, 1/2} \right)
   \, .
\end{eqnarray}
Apart from the so--called structure function $f_2^{3/2, 1/2}$
which describes the mixing between the leading electric--dipole
and the (much weaker) magnetic quadrupole decay channels and which
takes a value of about $f_2^{3/2, 1/2} = 1.28$ for the
hydrogen--like uranium U$^{91+}$ \cite{Surzhikovprl}, the angular
distribution (\ref{angular_Lyman}) depends on the five tensor
components $A^{(d)}_{2q}(\hat{n}_1; {\bm \epsilon}_{\chi_{_1}})$,
$q = -2,...,2$. As seen from Eqs.~(\ref{alignment_def}),
(\ref{amplitude_decomposition}) and (\ref{rhokq_dependence}) these
components are in general complex, and their imaginary and real
parts are related to each other as
$\mathrm{Re}(A^{(d)}_{21})=-\mathrm{Re}(A^{(d)}_{2-1})$,
$\mathrm{Im}(A^{(d)}_{21})=\mathrm{Im}(A^{(d)}_{2-1})$,
$\mathrm{Re}(A^{(d)}_{22})=\mathrm{Re}(A^{(d)}_{2-2})$,
$\mathrm{Im}(A^{(d)}_{22})=-\mathrm{Im}(A^{(d)}_{2-2})$.

\medskip

\begin{figure}
\begin{center}
\includegraphics[width=0.9\textwidth]{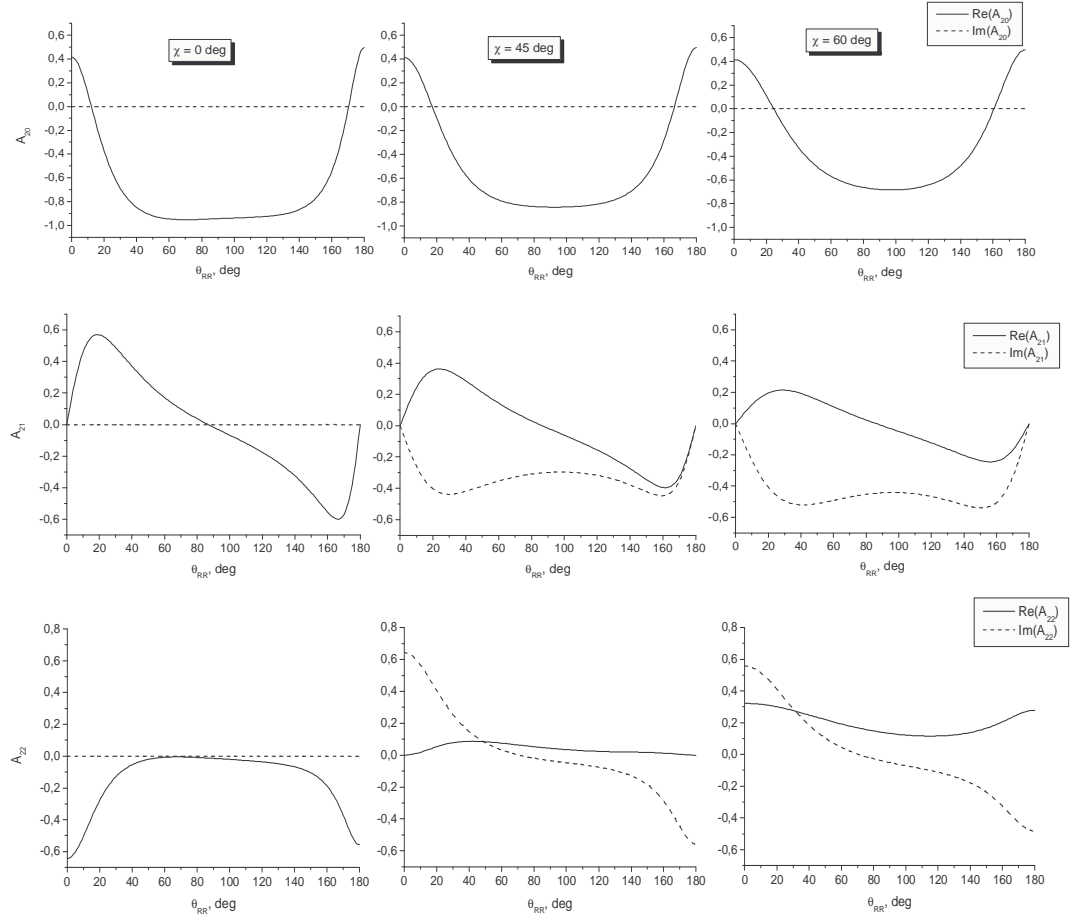}
\end{center}
\caption{The components $A_{20}$ (upper row), $A_{21}$ (middle row)
and $A_{22}$ (bottom row) of the reduced statistical tensor as
functions of the recombination photon emission angle $\theta_1
\equiv \theta_{RR}$. The real (solid line) and imaginary (dashed
line) parts of the components are displayed for the capture of an
unpolarized electron into the $2p_{3/2}$ state of a bare uranium ion
with projectile energy $T_p$ = 1 MeV/u, which correspond to the
kinetic energy $E=0.55$ keV of the incoming electron in the
rest--frame of the ion. Calculations are performed in the projectile
frame and for the emission of a recombination photon which is
linearly polarized under the angles $\chi = $0$^\circ$ (left
column), 45$^\circ$ (middle column) and 60$^\circ$ (right column)
with respect to the reaction plane.} \label{Fig5}
\end{figure}

Since the angular distribution (\ref{angular_Lyman}) of the
Lyman--$\alpha_1$ radiation is uniquely defined by the components
of the reduced statistical tensor $A^{(d)}_{2q}$, we shall first
investigate the dependence of these components on the emission
angle $\theta_1 \equiv \theta_{RR}$ and polarization ${\bm
\epsilon}_{\chi_{_1}}$ of the recombination photon. Figure
\ref{Fig5} displays the real and imaginary parts of the parameters
$A^{(d)}_{20}$, $A^{(d)}_{2 1}$ and $A^{(d)}_{2 2}$ for the
electron capture into the $2p_{3/2}$ state of initially bare
uranium ion with projectile energy $T_{p}=1 $MeV/u. Calculations
are performed in the projectile frame and for the emission of
\textit{linearly} polarized recombination photons with angles
$\chi_1$ = 0$^{\circ}$, 45$^{\circ}$ and 60$^{\circ}$ with regard
to the reaction plane. As expected, the tensor component
$A^{(d)}_{2 0}$ with zero projection $q=0$ is purely real for all
these polarization directions. In fact, this component represents
the differential (in angle and polarization) alignment parameter
and can be expressed in terms of the differential cross sections
as:
\begin{eqnarray}
   A^{(d)}_{2 0}(\hat{n}_1, \bm{\epsilon}_{\chi_{_1}}) =
   \frac{
   \frac{d\sigma_{\mu_d = +3/2}}{d\Omega_1}   +
   \frac{d\sigma_{\mu_d = -3/2}}{d\Omega_1}  -
   \frac{d\sigma_{\mu_d = +1/2}}{d\Omega_1}  -
   \frac{d\sigma_{\mu_d = -1/2}}{d\Omega_1}
   }
   {
   \frac{d\sigma_{\mu_d = + 3/2}}{d\Omega_1}  +
   \frac{d\sigma_{\mu_d = - 3/2}}{d\Omega_1}  +
   \frac{d\sigma_{\mu_d = + 1/2}}{d\Omega_1}  +
   \frac{d\sigma_{\mu_d = - 1/2}}{d\Omega_1}
   } \,
\end{eqnarray}
if the capture of the electron occurs into the magnetic substate
$\ketm{2p_{3/2} \mu_d}$ and under the simultaneous emission of a
photon with polarization vector $\bm{\epsilon}_{\chi_{_1}}$. As
seen from Fig.~\ref{Fig5}, the differential alignment parameter
$A^{(d)}_{2 0}$ is positive in the forward and backward
directions, referring to a preferred population of the two $\mu_d
= \pm 3/2$ substates. In contrast, the emission of a recombination
photon perpendicular to the beam mainly results in the population
of the $\mu_d = \pm 1/2$ substates which slowly varies from
97.2~\%{} for the polarization angle $\chi_1 = 0^\circ$ to
84.0~\%{} for $\chi_1 = 60^\circ$.

\medskip

Beside of the reduced statistical tensor $A^{(d)}_{2 0}$, which
refers to the differential alignment, the spin state of the
excited ions in the $2p_{3/2}$ state is also described by the
parameters $A^{(d)}_{2 \pm 1}$ and $A^{(d)}_{2 \pm 2}$, i.e.\ by
the non--diagonal elements of the density matrix. These
(additional) parameters also depend on the emission angle as well
as the polarization direction $\chi_1$ of the recombination
photon. While for the polarization angle $ \chi_1 = 0^\circ$ both
parameters $A^{(d)}_{2 \pm 1}$ and $A^{(d)}_{2 \pm 2}$ are again
purely real, they become complex when the linear polarization
vector of the recombination photon rotates out of the reaction
plane (cf.~Fig.~\ref{Fig5}).

\medskip

%
%
%

\begin{figure}
\epsfbox{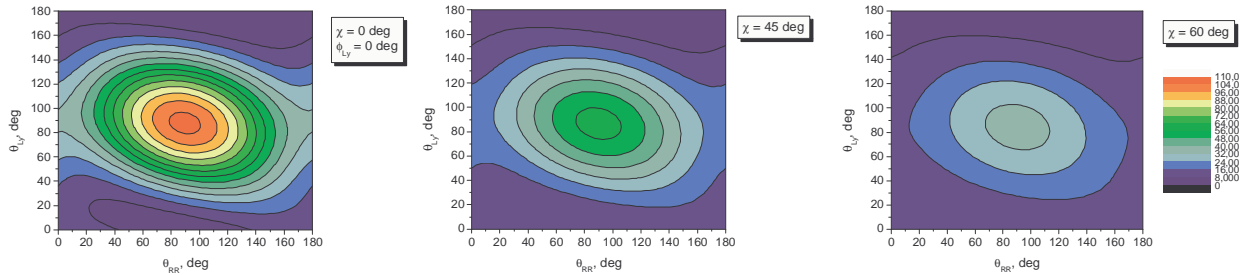}
\caption{ Differential RR cross section
(\ref{dcs_general_scenario_3}) as a function of the emission
angles of the recombination and the subsequent decay photons.
Calculations are performed within the ion--rest frame for the
projectile energy $T_p$ = 1 MeV/u, axial angle $\phi_{2} \equiv
\phi_{Ly} =$ 0$^\circ$ of the decay photon and for the emission of
the recombination photon being linearly polarized under the angles
$\chi = $0$^\circ$ (left panel), 45$^\circ$ (middle panel) and
60$^\circ$ (right panel) with respect to the reaction plane.}
\label{Fig6}
\end{figure}

\begin{figure}
\epsfbox{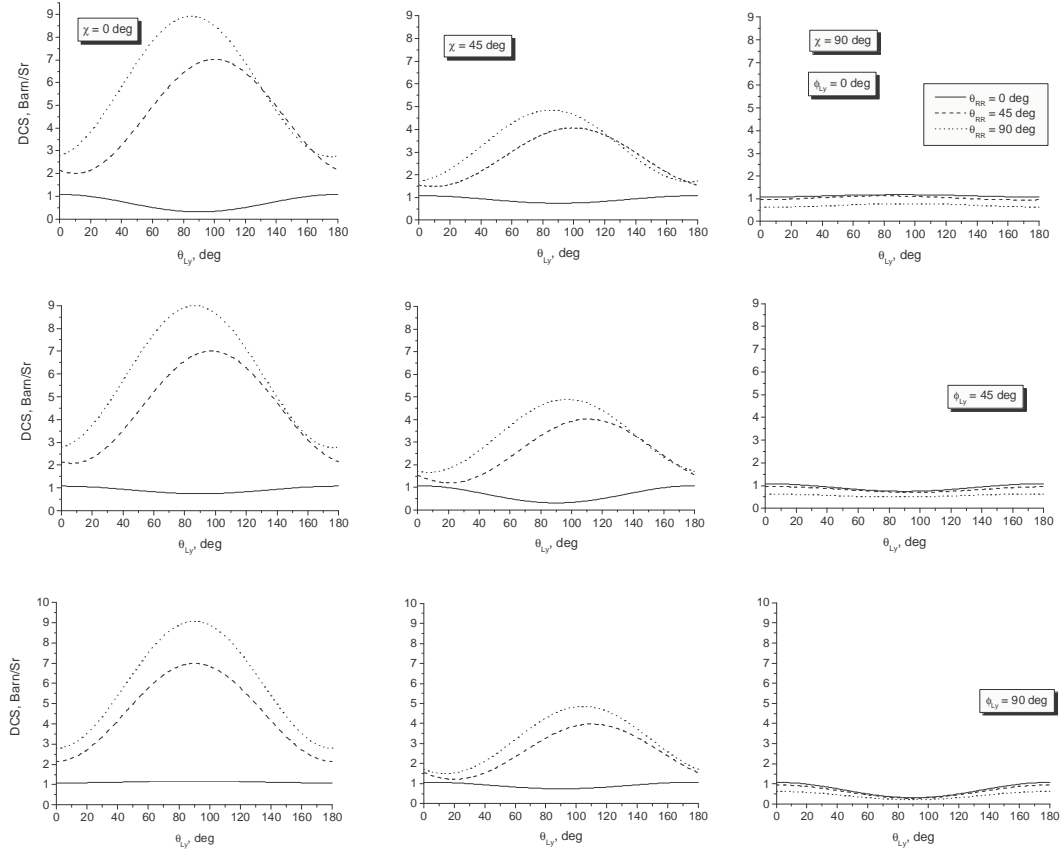} \caption{ Differential RR cross section
(\ref{dcs_general_scenario_3}) as a function of the emission
angles of the decay photons. Calculations are performed within the
ion--rest frame for the projectile energy $T_p$ = 1 MeV/u, axial
angle  $\phi_{Ly} = 0^\circ$ (upper row), 45$^\circ$ (middle row)
and 90$^\circ$ (lower row) of the decay photon, emission angle
$\theta_{RR} = 0^\circ$ (solid line), 45$^\circ$ (dashed line) and
90$^\circ$ (dotted line) of the recombination photon and for the
emission of the recombination photon being linearly polarized
under the angles $\chi = $0$^\circ$ (left column), 45$^\circ$
(middle column) and 90$^\circ$ (right column) with respect to the
reaction plane.} \label{Fig7}
\end{figure}

Having discussed the properties of the reduced statistical tensors
$A^{(d)}_{2 q}$, we are now prepared to study the angular
distribution (\ref{angular_Lyman}) of the subsequent
Lyman--$\alpha_1$ photons. In \textit{polarization--angle}
coincidence measurements, this distribution will depend on both,
the polarization state $\bm{\epsilon}_{\chi_{_1}}$ as well as the
angle under which the recombination photon is observed. In
Fig.~\ref{Fig6} and Fig.~\ref{Fig7}, for example, we display the
angle--differential cross section $d\sigma(\hat{n}_1,
\bm{\epsilon}_{\chi_{_1}}; \, \hat{n}_2) / d\Omega_2$ as
calculated for the RR of uranium projectile with energy $T_p =$ 1
MeV/u and for the emission of a linearly polarized recombination
photon with angles $\chi_1 =$ 0$^{\circ}$, 45$^\circ$ and
60$^\circ$, respectively. For these parameters we calculated the
differential cross section (\ref{angular_Lyman}) as a function of
the angles $\theta_1 \equiv \theta_{RR}$ and $\theta_2 \equiv
\theta_{Ly}$ of the recombination and the Lyman--$\alpha_1$
photon. Moreover, since the coincidence experiments, as planned at
the GSI storage ring, will be carried out most likely in a
coplanar geometry (that is, when both photons are detected within
the same plane), we have assumed here in the computations that
$\phi_{Ly}$ = 0$^{\circ}$. Again, as expected for this axial angle
and for a forward emission of the recombination photon
($\theta_{RR}$ = 0$^\circ$), the Lyman--$\alpha_1$ distribution is
symmetric around $\theta_{Ly}$ = 90$^\circ$ and also has its
minimum at this value since the differential alignment $A^{(d)}_{2
0}$ is positive in this case (cf.~Fig.~5) and the reduced tensor
components $A^{(d)}_{2 \pm 1}$ vanish identically. For all other
angles ($\theta_{RR} \ne$ 0$^\circ$ and $\theta_{RR} \ne$
180$^\circ$), these statistical tensors are generally non--zero
and give rise to an asymmetric distribution of the
Lyman--$\alpha_1$ photons in coincidence measurements. The
asymmetric shift in the angular distribution of the characteristic
radiation becomes most pronounced for the emission of the
recombination photons under the angle $\theta_{RR} = 17^{\circ}$
with respect to the beam direction.

%
%
%
%
%
%

%
%
%
%
%
%


\section{Summary}
\label{sec_summary}

In this paper, we re--investigated the radiative recombination of
a free electron into an excited state of a bare, high--Z ion and
its subsequent photon decay. Based on the resonant approximation
and the density matrix formalism we derived a general expression
for the double--differential RR cross section which accounts for
both, the angles and the polarization states of the recombination
and the subsequent decay photons. By making use of this
differential cross section we studied the polarization and angular
correlations between the two emitted photons. In particular, we
analyzed how the linear polarization of the characteristic
radiation depends on the particular angle under which the
recombination photon is observed. In a second scenario, the
correlations between the polarization states of the recombination
photons and the emission pattern of the subsequent decay have also
been discussed. Although the expressions derived in this paper can
be applied to any excited hydrogenic state, detailed calculations
were performed for the electron capture into the $2p_{3/2}$ state
of (initially) bare uranium ions U$^{92+}$ and its subsequent
Lyman--$\alpha_1$ ($2p_{3/2} \to 1s_{1/2}$) decay. These
calculations indicate a rather strong correlation between the
polarization states and emission patterns of the recombination and
decay photons. Apart from the coplanar geometry, which will be
utilized most likely by forthcoming experiments at the GSI storage
ring, detailed angular distributions of the Stokes parameters are
presented also for a non--coplanar set--up of the detectors. Such
a set--up is likely to be implemented at the Super--EBIT
facilities. In particular, the geometry of the Stockholm
University S--EBIT offers combinations of the observation angles
alternative to those available at the GSI storage ring.  The
angles $\theta_{RR}$ and $\theta_{Ly}$ are limited to 0$^{\circ}$
or 90$^{\circ}$ whereas the angle $\phi_{Ly}$ can be varied
between 0$^{\circ}$ and 360$^{\circ}$ in steps of 45$^{\circ}$.
The non coplanar geometry will be essential for the observation of
the linear polarization of the Lyman--$\alpha_1$ photons outside
of the reaction plane.


\section{Acknowledgments}
This work was supported by DFG (Grant No. 436RUS113/950/0-1) and
by RFBR (Grant No. 08-02-91967). The work of A.V.M. and V.M.S. is
also supported by the Ministry of Education and Science of Russian
Federation (Program for Development of Scientific Potential of
High School, Grant No. 2.1.1/1136). A.S. acknowledges support from
the Helmholtz Gemeinschaft (Nachwuchsgruppe VH--NG--421). S.F. is
grateful for the support by GSI under the project No.~KS--FRT.


\end{document}